\documentclass{article}
\usepackage[T1]{fontenc}
\usepackage[utf8]{inputenc}
\usepackage{ismir}

\makeatletter
\renewcommand{\permission}{}
\makeatother
\usepackage{amsmath,cite,url}
\usepackage{graphicx}
\usepackage{color}
\usepackage{algorithm}
\usepackage{algorithmic}
\usepackage{amsmath}

\usepackage{booktabs}

\title{RPPNet: Perceptually-Grouped Rhythm-Pitch Primitives for Long-Term Structure Melody Generation via Boundary-Aware Modeling} 

\multauthor
  {Tieyao Zhang$^1$ \hspace{1cm} Yuke Liu$^1$ \hspace{1cm} Jiaxing Yu$^2$}
  {Xinda Wu$^2$ \hspace{1cm} Kejun Zhang$^2$ \hspace{1cm} {\bf Genfang Chen$^{1,3}$}\\
  $^1$ Zhejiang Conservatory of Music, Hangzhou, China\\
  $^2$ Zhejiang University, Hangzhou, China\\
  $^3$ ZJU-ZJCM MusicAI Lab, Hangzhou, China\\
  {\tt\small tieyaozhang@stu.zjcm.edu.cn, keke@stu.zjcm.edu.cn, yujx@zju.edu.cn,} \\
  {\tt\small wuxinda@zju.edu.cn, zhangkejun@zju.edu.cn, cgf@zjcm.edu.cn}
  }

\def\authorname{T. Zhang, Y. Liu, J. Yu, X. Wu, K. Zhang, and G. Chen}

\begin{document}

\maketitle

\begin{abstract}
Existing symbolic music generation models typically use bars as the basic structural unit. However, human perception of musical phrases often does not align with notated bar lines, leading to long-term structural fragmentation. This paper proposes RPPNet—a two-stage deep learning architecture with variable structural boundaries. It first generates variable-length Rhythm-Pitch Primitive (RPP) sequences, where each RPP encodes note count, rhythm, and contour; then decodes the RPP sequences into concrete notes. The grouping of RPPs is automatically derived from acoustic cues, auditory inertia, and similarity perception based on music psychology. Experiments show that melodies generated by RPPNet are superior in both long-term structure and musicality, with significant improvements across all subjective evaluation dimensions. Ablation studies confirm that the performance gain stems from the structural correctness of the psychological representation, rather than from model capacity. This work offers an interdisciplinary perspective for music generation, integrating music theory, computational modeling, and music psychology.
\end{abstract}

\section{Introduction}\label{sec:introduction}
Symbolic melody generation is a core direction of generative AI in art \cite{jiSurveyDeepLearning2023}. In recent years, text‑to‑audio models have achieved notable results and can generate complete songs, yet they do not allow fine‑grained user control \cite{copetSimpleControllableMusic2023, liangWavCraftAudioEditing2024, liuAudioldmTexttoaudioGeneration2023}. Symbolic music, in contrast, enables precise control over melody, harmony, and other elements, holding a central position in human‑AI co‑creation. Consequently, this direction continues to attract significant attention from both academia and industry \cite{briotDeepLearningMusic2020}.

In recent years, researchers have drawn on language models from natural language processing to capture long‑term dependencies in music and achieve structured melody generation. Such methods offer an end‑to‑end framework, excellent representation learning ability, and the potential to generate sequences of arbitrary length. They require no hand‑crafted domain rules and can automatically extract knowledge from large‑scale data; their effectiveness has been validated across multiple disciplines \cite{annaMusicTransformerGenerating2018, jumperHighlyAccurateProtein2021, liNeuralSpeechSynthesis2019, schwallerExtractionOrganicChemistry2021, valanarasuMedicalTransformerGated2021}.

However, long-term structure modeling in symbolic music remains challenging \cite{carnovaliniComputationalCreativityMusic2020, maFoundationModelsMusic2024, yinDeepLearningsShallow2023}. Existing methods operate within linear sequence frameworks where attention mechanisms capture long-range token correlations yet fail to learn explicit associations between non-adjacent but functionally similar structural units; meanwhile, this task requires integrating music theory, computational modeling, psychology, and neuroscience \cite{baderSpringerHandbookSystematic2018}. Both factors jointly constrain effective learning of hierarchical logic, biasing models toward local statistical patterns and limiting long-term structure generation quality.

Recent works integrate computational modeling with music theory to improve long-term coherence through explicit structural boundaries and thematic features, such as Museformer  \cite{yuMuseformerTransformerFineand2022} with its hierarchical attention mechanism, Theme Transformer  \cite{shihThemeTransformerSymbolic2022} with thematic material retrieval, and MELONS \cite{zouMelonsGeneratingMelody2022} and PopMNet \cite{wuPopMNetGeneratingStructured2020} with graph-structured thematic variation encoding. However, these methods rely on fixed bar-level units. Psychological evidence shows that listeners perceive structural boundaries through acoustic cues, auditory inertia, and similarity-based grouping \cite{cooperRhythmicStructureMusic1963, deliegeGroupingConditionsListening1987, lamontMotivicStructurePerception2001}, rather than fixed metric segmentation. This mismatch hinders listeners' perceptual grouping of the generated structure, thereby reducing overall musical coherence.

To address this problem, we propose a hierarchical melody generation framework, RPPNet\footnote{The source code is  available at: \url{https://github.com/Kreuzter0421/RPPNet-PyTorch.git}}. It is based on an autoregressive architecture and splits generation into an RPP-level generation stage and a Note-level generation stage. Both stages use a Transformer encoder-decoder \cite{renPopmagPopMusic2020}.

Drawing on music theory and cognitive psychology, we use three perceptual dimensions—acoustic cues, auditory inertia, and similarity relations—and rhythmic structure analysis \cite{deliegeGroupingConditionsListening1987, edworthyIntervalContourMelody1985, edworthyPitchContourMusic1982} to design a heuristic algorithm. This algorithm represents a melody as a sequence of non-equidistant rhythm-pitch primitives (RPPs).

The RPP-level stage employs decoupled serial prediction for autoregressive RPP generation, constructing long-term structure; the Note-level stage decodes RPPs into MIDI notes via time-scale expansion mapping. Unlike bar-based approaches (e.g., Museformer, MELONS), RPPNet replaces fixed bars with perceptually-driven flexible grouping units—the first hierarchical model to do so, grounded in psychological evidence that phrase boundaries often misalign with notated bar lines.


Our contributions are as follows:
\begin{enumerate}
\item We propose RPPNet, a two-stage hierarchical framework with a novel perception-driven structural representation. It replaces fixed bar-level boundaries with flexible rhythm-pitch primitives grouped by acoustic cues, auditory inertia, and similarity perception.
\item We propose an automatic grouping algorithm that derives structural boundaries without explicit annotations. Controlled ablations demonstrate that performance gains stem from perceptual grouping principles rather than model capacity.
\item The proposed time-scale expansion mapping bridges discrete RPP events and continuous Note-level generation, with decoupled serial prediction ensuring strict alignment between generated notes and the RPP sequences.
\item Experimental results show that RPPNet outperforms existing baselines in long-term structural coherence and musicality.
\end{enumerate}
\section{Related Work}
\subsection{Long Sequence Symbolic Music Generation}
Symbolic music generation performs algorithmic composition by modeling discrete musical events, encoded in formats including MIDI, MusicXML, and ABC notation.Early research focused on local sequential dependency modeling: N-gram Markov chains predict the next event based on joint preceding events \cite{brooksExperimentMusicalComposition1957}, yet can only capture short-range dependencies and fail to model deep hierarchical structures. HMM introduce latent states to implicitly control token distributions  \cite{mavromatisHiddenMarkovModel2006, mavromatisHMMAnalysisMusical2009}, while dynamic Bayesian networks further explicitly model deep dependencies \cite{murphyDynamicBayesianNetworks2002}.

In recent years, deep neural network generative models have achieved significant advances in symbolic music. From RNNs \cite{chuSongPIMusically2017}, VAEs \cite{serbanHierarchicalLatentVariable2017}, and GANs \cite{yangMidiNetConvolutionalGenerative2017}, to Transformers \cite{daiTransformerxlAttentiveLanguage2019, katharopoulosTransformersAreRnns2020, vaswaniAttentionAllYou2017}, modeling capabilities have steadily increased. However, the quadratic complexity of Transformer self-attention, compounded by the long sequence nature of symbolic music, poses severe challenges for modeling dependencies spanning hundreds of bars.

This challenge stems from standard tokenization strategies, where a single piece can reach thousands to tens of thousands of tokens \cite{fradet2023miditok}, severely weakening long-term dependency modeling \cite{ouPhraseVAEPhraseLDMLatent2025, fradet2023byte}. Even when subsequent studies introduce compound tokens (e.g., Compound Word \cite{wuJazzTransformerFront2020}, Octuple \cite{zengMusicbertSymbolicMusic2021}) to compress sequence length, models still struggle to capture thematic repetition and variation spanning dozens of bars—the core difficulty of long-term structure modeling.
\subsection{Structure modeling guided by music theory}

To address the challenges of long‑range structural modeling in music, researchers have attempted to incorporate music theory priors into model design to guide globally coherent generation. The relevant methods can be grouped into three categories.

Structure‑aware attention: Optimizing attention distribution to efficiently extract musical structural features. For example, Museformer \cite{yuMuseformerTransformerFineand2022} combines fine‑ and coarse‑grained attention to capture bar‑level structure; HAT \cite{zhangStructureenhancedPopMusic2022} builds a harmony‑aware hierarchical Transformer, using harmonic progressions to guide long‑range dependency modeling.

External structural prior injection: Explicitly introducing boundary or thematic features to constrain the generation process. Ref. \cite{wuPowerFragmentationHierarchical2023} uses segmentation modules and multi‑scale attention to inject structural boundary information, aiding long‑form generation; Theme Transformer \cite{shihThemeTransformerSymbolic2022} takes retrieved thematic material as a conditional sequence to enhance cross‑segment thematic consistency.

Hierarchical modeling and task decomposition: break long sequence generation into multi-level subtasks for globally controllable structure. MusicFramework \cite{daiControllableDeepMelody2021} captures repetitive patterns via hierarchical representations to support multi-level abstract melody generation; TM-CGAN  \cite{jinThemeMusicGeneration2025} explicitly models themes with a VAE-CGAN hybrid architecture, balancing global coherence and local variation; PhraseLDM \cite{ouPhraseVAEPhraseLDMLatent2025} performs non-autoregressive generation in a phrase-level latent space to yield 128-bar multi-track music, alleviating long-term structure fragmentation. PopMNet \cite{wuPopMNetGeneratingStructured2020}  models internal dependencies via chord-melody graphs, while MELONS\cite{zouMelonsGeneratingMelody2022} combines bar-level relational graphs with GraphRNN and Transformer for long-term structure modeling.

However, these methods all lack perceptual grounding in structural segmentation. Neither explicit bar-level splitting nor implicit attention-based partitioning can capture the diversity and perceptual uncertainty of phrase boundaries, leading to structural degradation in long-term generation. Developing perceptually grounded, compact structural units for efficient structure-aware long-term structure modeling remains a core challenge.
\subsection{Perceptual Theories of Musical Structure}
The challenge of modeling long‑range structure in symbolic music \cite{carnovaliniComputationalCreativityMusic2020, maFoundationModelsMusic2024, yinDeepLearningsShallow2023} stems from the highly structured intrinsic nature of music: its elements are organized according to certain structural frameworks and expressed in human‑understandable ways \cite{baderSpringerHandbookSystematic2018}. This framework exhibits significant multi‑dimensional characteristics \cite{forteIntroductionSchenkerianAnalysis1982, koelschProcessingHierarchicalSyntactic2013, lerdahlGenerativeTheoryTonal1996}, encompassing not only basic components such as harmony and phrases but also interweaving cognitive organizational principles like similarity perception \cite{lamontMotivicStructurePerception2001}, making the analysis and modeling of its internal logic highly complex \cite{rohrmeierImplicitLearningAcquisition2012}.

Music psychology research shows that structural perception relies primarily on three mechanisms: (1) acoustic cues—using changes in intensity, pitch, duration, timbre, etc. to mark structural boundaries \cite{deliegeGroupingConditionsListening1987}; (2) auditory inertia—based on the establishment and fulfillment of expectations for musical events \cite{cooperRhythmicStructureMusic1963}; (3) similarity perception—associating structural elements through motivic repetition and variation \cite{lamontMotivicStructurePerception2001}. Together, these mechanisms reveal the hierarchical and theme‑driven nature of musical structure.

Deliege \cite{deliegeGroupingConditionsListening1987} validated the psychological effectiveness of grouping preference rules, confirming that acoustic changes in intensity and pitch serve as core cues for perceiving structural boundaries. Cooper and Meyer \cite{cooperRhythmicStructureMusic1963} identified auditory inertia as the organizing force of musical rhythm: listeners perceive continuous sound as a dynamic structure based on psychological expectations, where deviations such as syncopation and rests create tension. Lamont and Dibben \cite{lamontMotivicStructurePerception2001} demonstrated through cross-genre listening experiments that similarity judgments rely primarily on surface features such as texture and pitch contour rather than deep structural assumptions, exhibiting significant contextual dependence.

Traditional music modeling is mostly based on note‑ or bar‑level granularity, which hardly aligns with the actual mechanisms of structural perception. Drawing on the above psychological findings, this paper proposes a heuristic algorithm that automatically analyzes monophonic melody structure based on acoustic cues, auditory inertia, and similarity perception, and converts it into RPP sequence data.

\section{Method}\label{sec:page_size}
\subsection{Heuristic Segmentation of Rhythm-Pitch Primitive}
To construct cognitively grounded low-level structural units, we propose a heuristic melody segmentation algorithm that parses melodies into Rhythm-Pitch Primitives (RPPs), each containing at most three notes to carry local boundary information and support RPP-level generation, laying the foundation for subsequent hierarchical modeling.
\begin{figure*}[t] 
  \centering
  \includegraphics[width=0.95\textwidth]{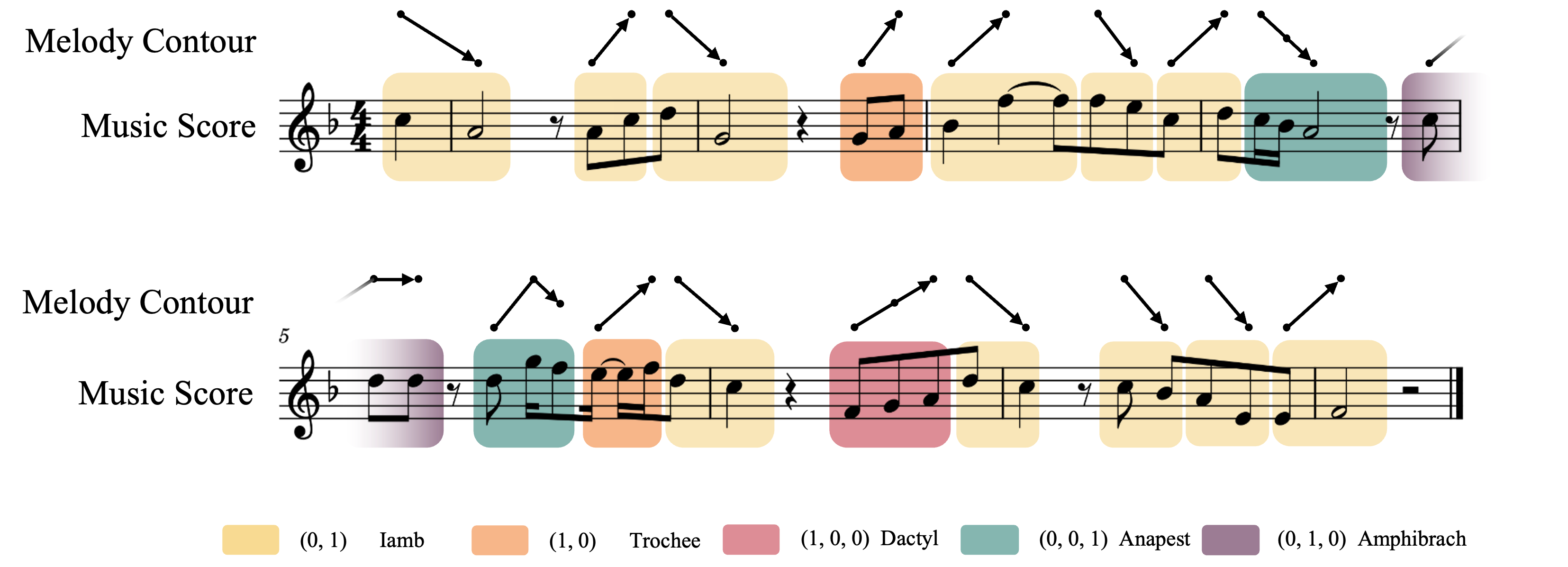} 
  \caption{Example of RPP grouping (excerpt from Hey Jude). Colored legends indicate rhythmic types; black arrows denote melodic contour direction.} 
  \label{fig:1}
\end{figure*}

Primitives are defined by rhythmic patterns (five categories, e.g., iambic) and pitch contours (ascending, descending, etc.), as illustrated in \figref{fig:1}; evolutionary relationships between adjacent primitives are encoded via multi-hot vectors. Grouping employs dynamic programming to maximize the cumulative structural weight $W(\cdot)$\footnote{The complete formalization of $W(\cdot)$ is available in the repository mentioned above.} of the parsed sequence (Algorithm \ref{alg:rpp_dp}), where $W(\cdot)$ scores each note based on its metrical grid position, duration, and syncopation pattern.
\begin{algorithm}[t] 
\caption{Dynamic Programming for RPP Grouping}
\label{alg:rpp_dp}
\begin{algorithmic}[1]
\REQUIRE Note sequence $P_{1:L}$, Weight function $W(\cdot)$
\ENSURE Optimal RPP list $\mathcal{R}$

\IF{$L \leq 3$}
    \RETURN $\{\text{MakeRPP}(P_{1:L})\}$
\ENDIF

\STATE $dp[0] \gets 0$, \quad $dp[1 \dots L] \gets -\infty$
\STATE $path[0 \dots L] \gets \emptyset$

\FOR{$i = 1$ \TO $L$}
    \STATE $k^* \gets \arg\max_{1 \le k \le 3} \{ dp[i-k] + W(P_{i-k+1:i}) \}$
    \STATE $dp[i] \gets dp[i-k^*] + W(P_{i-k^*+1:i})$
    \STATE $path[i] \gets path[i-k^*] \cup \{k^*\}$
\ENDFOR

\STATE $\mathcal{R} \gets \emptyset$, \quad $idx \gets 1$
\FOR{$k \in path[L]$}
    \STATE $\mathcal{R} \gets \mathcal{R} \cup \{\text{MakeRPP}(P_{idx:idx+k-1})\}$
    \STATE $idx \gets idx + k$
\ENDFOR

\RETURN $\mathcal{R}$
\end{algorithmic}
\end{algorithm}
\subsection{Hierarchical Sequence Representation}
To overcome self-attention dilution in long sequences \cite{yuMuseformerTransformerFineand2022}, we propose a hierarchical generation architecture that strictly decouples the RPP-level and Note-level.

\subsubsection{RPP-Level Representation}
We model the global structure as a sequence of Rhythm-Pitch Primitives. The sequence is $\mathcal{V} = \{v_1, \dots, v_T\}$. Each unit is explicitly defined as $v_i = \{b_i, p_i, d_i, r_i, m_i\}$, where $b_i$, $p_i$, and $d_i$ denote the measure index, metrical position, and duration respectively. The rhythmic pattern $r_i$ and melodic contour $m_i$ characterize the motif. This abstraction achieves significant temporal downsampling, enabling efficient identification of form development without external priors.

\subsubsection{Note-Level Representation}
To realize the RPP-level representation as Note-level outputs, we define a note sequence $\mathcal{U} = \{u_1, \dots, u_K\}$ mapping to final MIDI events. Each unit $u_j = \{b_j, p_j, d_j, n_j\}$ shares the aforementioned temporal grid, while $n_j$ explicitly defines the absolute pitch.

\subsubsection{Compound Word Embedding Strategy}
Extending the Compound Word parallel fusion concept \cite{hsiaoCompoundWordTransformer2021}, the discrete attributes within $v_i$ and $u_j$ are independently mapped into dense vectors at each time step. These vectors are subsequently aggregated into a unified hidden representation via concatenation and linear projection.
\subsection{Model}
\begin{figure}[t]
  \centering
  \includegraphics[width=0.9\linewidth]{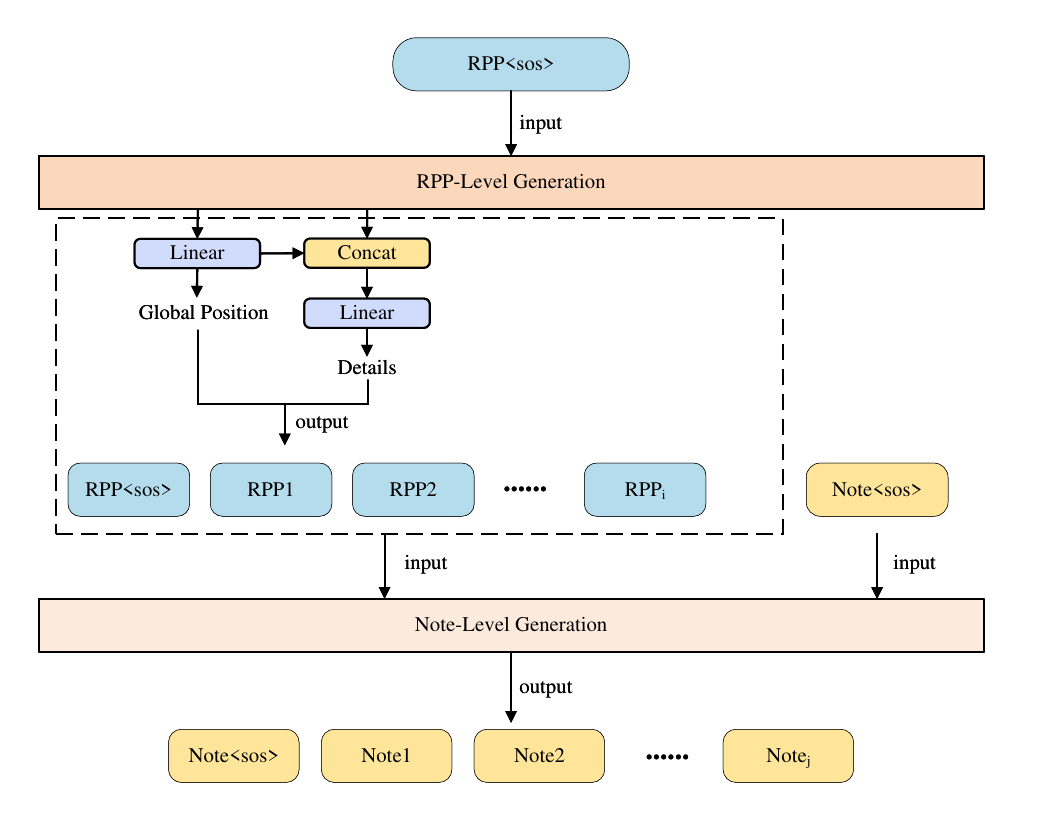} 
  \caption{The architecture of the proposed two-stage hierarchical generation framework. Melody construction is strictly decoupled into RPP-level structural planning and Note-level note instantiation. The dashed box details the decoupled serial prediction mechanism deployed at the output end.} 
  \label{fig:2}
\end{figure}

As shown in \figref{fig:2}, our two-stage hierarchical generation framework decouples melody construction into RPP-level structural planning and Note-level instantiation. This multi-resolution temporal representation effectively addresses the semantic disparity between abstract structural development and Note-level symbolic generation.

\subsubsection{RPP-Level Generation}
The RPP-level structural module aims to generate a long-term framework guiding musical development, featuring a standard Transformer encoder-decoder architecture as its core \cite{vaswaniAttentionAllYou2017}.

While employing compound representation inputs to compress sequence length, we replace traditional parallel prediction with a decoupled serial prediction mechanism at the output, highlighted by the dashed box in \figref{fig:2}. Given historical context features $\mathbf{h}_{<t}$ at time step $t$, the joint probability of primitive $v_t$ is strictly decoupled into the marginal probability of temporal position $pos_t$ and conditional probabilities of detailed attributes, formulated in \eqnref{eq:decoupled_prob}:
\begin{equation}\label{eq:decoupled_prob}
P(v_t | \mathbf{h}_{<t}) = P(pos_t | \mathbf{h}_{<t}) \cdot P(d_t, r_t, m_t| \mathbf{h}_{<t}, pos_t)
\end{equation}
The model first predicts $pos_t$ solely from $\mathbf{h}_{<t}$. This position representation is subsequently mapped into a temporal prior embedding and concatenated with $\mathbf{h}_{<t}$, enabling the joint prediction of duration, rhythmic pattern, and melodic contour under explicit temporal constraints.



\subsubsection{Note-Level Generation}
The Note-level module decodes the RPP sequence into concrete note events via a Transformer encoder-decoder \cite{vaswaniAttentionAllYou2017}. We construct a time-scale expansion mapping that bridges discrete RPP events and the continuous note-level time axis: for each primitive $i$, the absolute onset is computed as $T_{start} = b_i \times \mathcal{R}_{bar} + p_i$, and its hidden state $\mathbf{h}_i$ is broadcast across all grid steps within $[T_{start}, T_{start} + d_i)$. This allows the decoder to attend to structural memory via cross-attention, ensuring strict alignment between output notes and the RPP-level skeleton.

\section{Experiment}\label{sec:typeset_text}
\subsection{Dataset}\label{subsec:body}
We adopt the MelodyNet dataset  \cite{wuMelodyGLMMultitaskPretraining2023}, comprising over 300,000 MIDI melodies from FreeMIDI, HookTheory, BitMIDI, MuseScore, KernScores, and Kunstderfuge. After removing non-melody tracks and incomplete data, 274,300 melodies remain, split into train/validation/test sets at a ratio of $18:1:1$.
\subsection{Compared Models}
To evaluate the structural awareness of RPPNet, we select two representative baseline models, human-composed melodies, and two framework variants to validate the specific contribution of the proposed heuristic RPP grouping:
\begin{itemize}
\item Museformer \cite{yuMuseformerTransformerFineand2022}: A long-sequence music generation model with fine- and coarse-grained attention.
\item MELONS \cite{zouMelonsGeneratingMelody2022}: A graph-sequence architecture that models bar-level chord and melodic contours via line graphs to condition token generation.
\item RPPNet-Real: A variant using ground-truth RPP sequences as RPP-level input, representing the theoretical ceiling of Note-level generation.
\item RPPNet-Random-Grouped: An ablation baseline replacing heuristic RPP grouping with random segmentation while maintaining architectural consistency and length distribution alignment.
\end{itemize}
We select Museformer \cite{yuMuseformerTransformerFineand2022} and MELONS \cite{zouMelonsGeneratingMelody2022} as primary baselines, as both exemplify fixed bar-level structural modeling — the exact assumption our work relaxes. MelodyGLM \cite{wuMelodyGLMMultitaskPretraining2023} is excluded because its pretrain-finetune paradigm confounds structural boundary design with large-scale pretraining. PhraseLDM \cite{ouPhraseVAEPhraseLDMLatent2025} is excluded as it requires explicit phrase-level annotations unavailable in MelodyNet. Our ablation baseline RPPNet-Random-Grouped isolates the contribution of perceptual semantics from mere variable-length grouping, yielding a clean controlled comparison: fixed-bar (Museformer, MELONS) vs. perceptually-grouped variable-length (RPPNet) vs. randomly-grouped variable-length (ablation).
\subsection{Experiment Setup}
\subsubsection{Objective Metrics}
To validate the effectiveness of RPP grouping, we construct a random-grouping baseline (Random-Grouped) that preserves the length distribution of the RPP training set via random sampling while removing structural semantics.

Following Museformer \cite{yuMuseformerTransformerFineand2022}, we adopt Perplexity (PPL) and Structural Error (SE) as objective metrics. This alignment isolates the effect of structural representation choice from metric selection bias. PPL requires a consistent vocabulary across models, so Museformer and MELONS are excluded from the objective comparison. We also note growing evidence questioning the perceptual validity of conventional symbolic music objective metrics \cite{SurveyontheEvaluation2025}.

We further verify the RPP-level generator's learning of underlying distributions by comparing attribute statistics between ground-truth and generated RPP sequences.
\subsubsection{Subjective Metrics}
Following \cite{wuMelodyGLMMultitaskPretraining2023}, we conduct a subjective listening test with 15 paid participants (7 with musical background, 8 without) rating generated melodies on a 10-point scale (1–-10). All models are trained from scratch, yielding 50 32-bar melodies evaluated across five configurations—RPPNet, RPPNet-Real, RPPNet-Random-Grouped, Museformer, and MELONS—alongside 10 human-composed references from the test set.

Participants were asked to rate each melody on a scale of 1--10 (lowest to highest) along the following four dimensions:
\begin{itemize}
\item Coherence: Is the melody fluent, pleasant, and engaging?
\item Rhythmicity: Does the melody exhibit a regular metric pattern and appropriate use of rests?
\item Structure: Does the melody display clear structural features, such as reasonable repetition and motivic development?
\item Overall Impression: A holistic evaluation of the melody.
\end{itemize}

We select Museformer \cite{yuMuseformerTransformerFineand2022} and MELONS  \cite{zouMelonsGeneratingMelody2022}  as primary baselines, both representing fixed bar-level structural modeling that our work relaxes. MelodyGLM \cite{wuMelodyGLMMultitaskPretraining2023} is excluded as its pretrain-finetune paradigm introduces a confounding variable unrelated to boundary design. PhraseLDM \cite{ouPhraseVAEPhraseLDMLatent2025} is excluded due to its reliance on explicit phrase annotations unavailable in MelodyNet. The ablation baseline RPPNet-Random-Grouped isolates perceptual semantics from variable-length grouping alone, enabling a controlled comparison across fixed-bar, perceptually-grouped, and randomly-grouped conditions.

\section{Result}
\begin{table*}[t]
  \centering
  \small
  \begin{tabular}{lllll}
    \toprule
    \textbf{Model} & \textbf{Coherence} & \textbf{Rhythmicity} & \textbf{Structure} & \textbf{Overall} \\ 
    \midrule
    Human & 7.29 $\pm$ 0.74$^{**}$ & 7.36 $\pm$ 0.68$^{**}$ & 7.45 $\pm$ 0.82$^{**}$ & 7.47 $\pm$ 0.79$^{**}$ \\
    \midrule
    \textbf{RPPNet (Ours)} & 6.61 $\pm$ 0.71 & 6.61 $\pm$ 0.69 & 6.77 $\pm$ 0.73 & 6.69 $\pm$ 0.68 \\
    RPPNet-Real & 6.61 $\pm$ 0.85 & 6.59 $\pm$ 0.83 & 6.51 $\pm$ 0.98 & 6.61 $\pm$ 0.87 \\
    RPPNet-Rnd-Grp & 5.99 $\pm$ 0.62$^{***}$ & 6.08 $\pm$ 0.54$^{**}$ & 6.01 $\pm$ 0.64$^{***}$ & 6.09 $\pm$ 0.66$^{**}$ \\
    Museformer \cite{yuMuseformerTransformerFineand2022} & 5.81 $\pm$ 0.54$^{***}$ & 5.62 $\pm$ 0.53$^{***}$ & 5.57 $\pm$ 0.58$^{***}$ & 5.73 $\pm$ 0.54$^{***}$ \\
    MELONS \cite{zouMelonsGeneratingMelody2022} & 5.39 $\pm$ 0.69$^{***}$ & 5.33 $\pm$ 0.66$^{***}$ & 5.31 $\pm$ 0.76$^{***}$ & 5.30 $\pm$ 0.77$^{***}$ \\
    \bottomrule
  \end{tabular}
  \caption{Results of the subjective evaluation. Scores are presented as Mean $\pm$ Standard Deviation. Asterisks denote significant differences compared to our proposed RPPNet ($^{**} p < 0.01$, $^{***} p < 0.001$). ``Overall'' denotes the overall musical impression.}
  \label{tab:subjective_eval}
\end{table*}
\begin{table}[t]
  \centering
  \small
  \begin{tabular}{lcc}
    \toprule
    \textbf{Attribute / Pattern} & \textbf{Real (\%)} & \textbf{Generated (\%)} \\ 
    \midrule
    \textit{Note Group} & & \\
    One note           & 12.26 & 10.61 \\
    Two notes          & 36.25 & 40.41 \\
    Three notes          & 51.49 & 48.98 \\
    \midrule
    \textit{Rhythm Pattern} & & \\
    (0,)                  & 7.74  & 8.43  \\
    (1,)                  & 4.51  & 2.19  \\
    (0, 1)                & 14.29 & 29.83 \\
    (1, 0)                & 21.96 & 10.57 \\
    (0, 0, 1)             & 14.14 & 16.31 \\
    (0, 1, 0)             & 17.72 & 17.18 \\
    (1, 0, 0)             & 19.64 & 15.49 \\
    \midrule
    \textit{Melodic Contour} & & \\
    Descending            & 15.47 & 15.15 \\
    Ascending             & 13.18 & 18.19 \\
    Stationary            & 7.60  & 7.06  \\
    Cont. Descending      & 7.68  & 6.83  \\
    Cont. Ascending       & 7.26  & 6.28  \\
    Cont. Stationary      & 6.32  & 3.25  \\
    Down then Up          & 8.93  & 8.91  \\
    Up then Down          & 10.11 & 8.15  \\
    Stay then Up          & 2.85  & 4.46  \\
    Stay then Down        & 3.30  & 3.98  \\
    Down then Stay        & 2.51  & 3.74  \\
    Up then Stay          & 2.54  & 3.39  \\
    Single Note           & 12.26 & 10.61 \\
    \bottomrule
  \end{tabular}
  \caption{Detailed statistical distributions of note density, rhythm pattern, and melodic contour across real and generated datasets.}
  \label{tab:comprehensive_music_stats}
\end{table}

\begin{table}[t]
  \centering
  \small
  \begin{tabular}{lcc}
    \toprule
    \textbf{Model} & \textbf{PPL} $\downarrow$ & \textbf{SE} $\downarrow$ \\ 
    \midrule
    RPPNet (Ours) & 2.21 & 0.0132 $\pm$ 0.0013 \\
    RPPNet-Random-Grouped & 2.36 & 0.0175 $\pm$ 0.0008 \\
    \bottomrule
  \end{tabular}
  \caption{Results of the objective evaluation in the ablation study. SE denotes the structural entropy of Rhythmic Pattern Primitives (RPPs). The best results are highlighted in bold.}
  \label{tab:ablation_study}
\end{table}
\subsection{Statistic Analysis}
To quantify RPP-level distributional fidelity, we statistically compare real and generated attribute distributions, yielding a Pearson correlation of $92.19\%$ across 23 categories (\tabref{tab:comprehensive_music_stats}). Dimension-level analysis reveals strong alignment in note group ($r = 0.98$) and melodic contour ($r = 0.90$), while rhythm pattern shows the largest gap ($r = 0.44$), primarily due to over-generation of iambs (0,1: $+15.5\%$) and under-generation of trochees (1,0: $-11.4\%$). These results suggest the model effectively captures note grouping and pitch contour structure, while rhythmic pattern fidelity remains the primary direction for improvement.
\subsection{Ablation Study}
To validate the effectiveness of heuristic RPP grouping, we construct a random-grouping ablation baseline (Random-Grouped) within the RPPNet framework with aligned length distribution. Objective metrics (\tabref{tab:ablation_study}) show that standard RPPNet ($PPL = 2.21$, $SE = 0.0132$) outperforms Random-Grouped (2.36, 0.0175).

Subjective evaluation (\tabref{tab:subjective_eval}) further reveals that Random-Grouped exhibits significant degradation across all dimensions upon removing heuristic grouping: coherence (5.99, p<0.001), rhythmic quality (6.08, $p<0.01$), structural quality (6.01, $p<0.001$), and overall impression (6.09, $p<0.01$). This demonstrates that the structural semantics captured by heuristic grouping are critical for melodic coherence and motif organization.

Meanwhile, standard RPPNet and RPPNet-Real (ground-truth RPP input) show no significant difference in coherence (6.61 vs. 6.61, $p=1.00$), rhythmic quality (6.61 vs. 6.59, $p=0.93$), or overall impression (6.69 vs. 6.61, $p=0.55$), with structural quality also approaching non-significance (6.77 vs. 6.51, $p=0.08$). This suggests that the RPP-level generator of RPPNet possesses strong distribution fitting capabilities, corroborating the findings from statistical experiments.

\subsection{Comparison with Previous Models}
\tabref{tab:subjective_eval} presents the subjective ratings and significance tests against baselines. Human-composed melodies score 7.47 overall as the reference ceiling. RPPNet achieves an overall impression of 6.69, significantly outperforming Museformer (5.73, $p<0.001$) and Melons (5.30, $p<0.001$). The advantage is most pronounced in structural quality: RPPNet (6.77) exceeds Museformer (5.57, $p<0.001$) and MELONS (5.31, $p<0.001$) by 1.20 and 1.46 points, respectively, demonstrating that explicit hierarchical structural modeling effectively enhances the organization of long-term structure.

\section{Discussion}
This work integrates music-psychological principles with music theory to propose RPPNet, a perception-driven structural representation for melody generation that transcends fixed bar-level boundaries through hierarchical decoupled generation. Experiments confirm that this framework significantly outperforms existing baselines in long-term structural coherence and overall musicality.

These results demonstrate that perception-driven grouping significantly outperforms fixed bar-level representations. Ablation experiments reveal that performance gains stem from the proper encoding of perceptual grouping principles—merely breaking bar isometry without respecting perceptual rules yields no benefit.

Although RPPNet significantly outperforms existing baselines, a quality gap remains compared to human compositions (overall $7.47 vs. 6.69$), likely attributable to the distributional mismatch in rhythmic patterns (Pearson $r = 0.44$). Closing this gap is the central challenge going forward.

Future work will proceed along three directions: (1) hybrid data-driven and rule-driven structural strategies; (2) introducing micro-macro feedback mechanisms where Note-level generation revises RPP-level structure; and (3) extending this representation to polyphonic music and accompaniment generation, evaluating its capability under complex musical textures.

\bibliography{RPPNet}

%
%
%
%

\end{document}